\documentclass[equations,11pt]{article}

\renewcommand{\theequation}{\arabic{equation}}
\usepackage{amssymb,amsmath}
\usepackage{color}
\usepackage{accents}
\usepackage{eucal}
\usepackage{slashed}
\usepackage{acronym}
\usepackage{amsfonts}
\usepackage{amsthm}
\usepackage{mathdots}
\usepackage{mathrsfs}
\usepackage{mathtools}
\usepackage{cancel}
\usepackage{todonotes}
\usepackage{setspace}

\newcommand{\mathsym}[1]{{}}
\newcommand{\unicode}[1]{{}}

\begin{document}
\bibliographystyle{plain}
\def\m@th{\mathsurround=0pt}
\mathchardef\bracell="0365
\def\upbrall{$\m@th\bracell$}
\def\undertilde#1{\mathop{\vtop{\ialign{##\crcr
    $\hfil\displaystyle{#1}\hfil$\crcr
     \noalign
     {\kern1.5pt\nointerlineskip}
     \upbrall\crcr\noalign{\kern1pt
   }}}}\limits}
\def\theequation{\arabic{section}.\arabic{equation}}
\newcommand{\ar}{\alpha}
\newcommand{\al}{\alpha}
\newcommand{\aar}{\bar{a}}
\newcommand{\bb}{\beta}
\newcommand{\gm}{\gamma}
\newcommand{\Gm}{\Gamma}
\newcommand{\en}{\ven}
\newcommand{\ven}{\varepsilon}
\newcommand{\dd}{\delta}
\newcommand{\sg}{\sigma}
\newcommand{\kp}{\kappa}
\newcommand{\ld}{\lambda}
\newcommand{\oa}{\omega}
\newcommand{\be}{\begin{equation}}
\newcommand{\ee}{\end{equation}}
\newcommand{\bag}{\begin{align}}
\newcommand{\eag}{\end{align}}
\newcommand{\bea}{\begin{eqnarray}}
\newcommand{\eea}{\end{eqnarray}}
\newcommand{\bse}{\begin{subequations}}
\newcommand{\ese}{\end{subequations}}
\newcommand{\nn}{\nonumber}
\newcommand{\bR}{\bar{R}}
\newcommand{\bP}{\bar{\Phi}}
\newcommand{\bS}{\bar{S}}
\newcommand{\bW}{\bar{W}}
\newcommand{\vf}{\varphi}
\newcommand{\sn}{{\rm sn}}
\newcommand{\wh}{\widehat}
\newcommand{\wb}{\overline}
\newcommand{\wt}{\widetilde}
\newcommand{\ut}{\undertilde}
\newcommand{\ub}[1]{\underline{#1}}
\newcommand{\uh}{\underhat}
\def \c#1{\accentset{\circ}{#1}}
\def \th#1{\widehat{\widetilde{#1}}}
\def \tb#1{\overline{\widetilde{#1}}}
\def \hb#1{{\widehat{\overline{#1}}}}
\def \dh#1{\underaccent{\hat}{#1}}
\def \dt#1{\underaccent{\tilde}{#1}}
\def \t#1{\widetilde{#1}}

\newcommand{\ip}{{i^\prime}}
\newcommand{\jp}{{j^\prime}}
\newcommand{\cn}{{\rm cn}}
\newcommand{\dn}{{\rm dn}}
\newcommand{\bun}{\boldsymbol{1}}
\newcommand{\Ld}{{\boldsymbol \Lambda}}
\newcommand{\tLd}{\,^{t\!}{\boldsymbol \Lambda}}
\newcommand{\tL}{\,^{t\!}{\mathcal{L}}}
\newcommand{\bh}{\boldsymbol{h}}
\newcommand{\bee}{\boldsymbol{e}}
\newcommand{\tee}{\,^t\!\boldsymbol{e}}
\newcommand{\bI}{{\boldsymbol I}}
\newcommand{\bO}{{\boldsymbol O}}
\newcommand{\bU}{{\boldsymbol U}}
\newcommand{\bC}{{\boldsymbol C}}
\newcommand{\bOm}{{\boldsymbol \Omega}}
\newcommand{\buk}{{\boldsymbol u}_\kappa}
\newcommand{\bck}{{\boldsymbol c}_\kappa}
\newcommand{\bul}{{\boldsymbol u}_\ell}
\newcommand{\tII}{\,^{t\!}{\boldsymbol I}}
\newcommand{\tvk}{\,^{t\!}{\boldsymbol v}_{\kappa^\prime}}
\newcommand{\thh}{\,^{t\!}h}
\newcommand{\tvl}{\,^{t\!}{\boldsymbol v}_{\ell^\prime}}
\newcommand{\tvll}{\,^{t\!}{\boldsymbol v}_{-\ell+\ld}}
\newcommand{\tuq}{\,^{t\!}{\boldsymbol u}_{-q_j+\ld}}
\newcommand{\tcl}{\,^{t\!}{\boldsymbol c}_{\ell}}
\newcommand{\tclp}{\,^{t\!}{\boldsymbol c}_{\ell^\prime}}
\newcommand{\tck}{\,^{t\!}{\boldsymbol c}_{\kp^\prime}}
\newcommand{\ssk}{\sigma_{\kappa^\prime}}
\newcommand{\ssl}{\sigma_{\ell^\prime}}
\newcommand{\pte}{(\partial_t-\partial_\eta)}
\newcommand{\pxe}{(\partial_x-\partial_\eta)}
\newcommand{\dint}{\int_\Gamma {\rm d}\mu(\ell) }

\newcommand{\ddint}{\iint_{D} {\rm d}\mu(\ell,\ell^\prime)}

\def\hypotilde#1#2{\vrule depth #1 pt width 0pt{\smash{{\mathop{#2}
\limits_{\displaystyle\widetilde{}}}}}}
\def\hypohat#1#2{\vrule depth #1 pt width 0pt{\smash{{\mathop{#2}
\limits_{\displaystyle\widehat{}}}}}}
\def\hypo#1#2{\vrule depth #1 pt width 0pt{\smash{{\mathop{#2}
\limits_{\displaystyle{}}}}}}
\def\hypobar#1#2{\vrule depth #1 pt width 0pt{\smash{{\mathop{#2}
\limits_{\displaystyle\bar{}}}}}}

\newcommand{\bblu}{\begin{color}{blue}}
\newcommand{\bred}{\begin{color}{red}}
\newcommand{\bgreen}{\begin{color}{green}}
\newcommand{\ecl}{\end{color}}

\newcommand{\Ups}{\Upsilon}
\newcommand{\bA}{\boldsymbol{A}}
\newcommand{\bB}{\boldsymbol{B}}
\newcommand{\bD}{\boldsymbol{D}}
\newcommand{\bE}{\boldsymbol{E}}
\newcommand{\bF}{\boldsymbol{F}}
\newcommand{\bG}{\boldsymbol{G}}
\newcommand{\bH}{\boldsymbol{H}}
\newcommand{\bJ}{\boldsymbol{J}}
\newcommand{\bK}{\boldsymbol{K}}
\newcommand{\bL}{\boldsymbol{L}}
\newcommand{\bM}{\boldsymbol{M}}
\newcommand{\bN}{\boldsymbol{N}}
\newcommand{\bQ}{\boldsymbol{Q}}
\newcommand{\bbR}{\boldsymbol{R}}
\newcommand{\bbS}{\boldsymbol{S}}
\newcommand{\bT}{\boldsymbol{T}}
\newcommand{\bV}{\boldsymbol{V}}
\newcommand{\bX}{\boldsymbol{X}}
\newcommand{\bsY}{\boldsymbol{Y}}
\newcommand{\bZ}{\boldsymbol{Z}}

\newcommand{\br}{\boldsymbol{r}}

\newcommand{\mbe}{{\boldsymbol e}}
\newcommand{\pl}{\partial}
\newcommand{\bnab}{{\boldsymbol \nabla}}
\newcommand{\bu}{\boldsymbol{u}}
\newcommand{\bv}{{\boldsymbol v}}
\newcommand{\ba}{{\boldsymbol a}}
\newcommand{\bbb}{\boldsymbol{b}}
\newcommand{\bc}{\boldsymbol{c}}
\newcommand{\bd}{\boldsymbol{d}}
\newcommand{\bme}{\boldsymbol{e}}
\newcommand{\bff}{\boldsymbol{f}}
\newcommand{\bk}{\boldsymbol{k}}
\newcommand{\bl}{\boldsymbol{l}}
\newcommand{\brr}{\boldsymbol{r}}
\newcommand{\bw}{{\boldsymbol w}}
\newcommand{\mbx}{{\boldsymbol x}}
\newcommand{\mby}{{\boldsymbol y}}
\newcommand{\bz}{{\boldsymbol z}}
\newcommand{\bp}{{\boldsymbol p}}
\newcommand{\bs}{{\boldsymbol s}}
\newcommand{\btt}{{\boldsymbol t}}
\newcommand{\bmm}{{\boldsymbol m}}
\newcommand{\bdd}{{\boldsymbol \delta}}
\newcommand{\bze}{{\boldsymbol 0}}
\newcommand{\boma}{{\boldsymbol \omega}}
\newcommand{\bet}{{\boldsymbol \eta}}
\newcommand{\bphi}{{\boldsymbol \phi}}
\newcommand{\bpsi}{{\boldsymbol \psi}}
\newcommand{\bkp}{{\boldsymbol \kappa}}
\newcommand{\bxi}{{\boldsymbol \xi}}
 \newcommand{\mbv}{\boldmath{v}}
 \newcommand{\mbxi}{\boldmath{\xi}}
 \newcommand{\mbeta}{\boldmath{\eta}}
 \newcommand{\mbw}{\boldmath{w}}
 \newcommand{\mbu}{\boldmath{u}}
\newcommand{\tbphi}{\,^{t\!}\boldsymbol{\bphi}}

\newcommand{\ts}{\,^t \hskip -2pt {\boldsymbol{s}}}
\newcommand{\tv}{\,^t \hskip -2pt {\boldsymbol{v}}}
\def \tyb#1{\hbox{\tiny{[{\it{#1}}]}}}
\def \ty#1{\hbox{\tiny{{\it{#1}}}}}


\begin{center}
{\Large{\bf Elliptic solutions of the lattice CKP equation and its elliptic direct linearisation scheme} } \\
\vspace{.2cm}
\emph{Ying-ying Sun\,\footnote{E-mail: yingying.sun@usst.edu.cn}, Da-jun Zhang\,\footnote{ E-mail: djzhang@staff.shu.edu.cn} and Frank Nijhoff\, \footnote{ E-mail: f.w.nijhoff@leeds.ac.uk}}

\it{\small 
1. Department of Mathematics, University of Shanghai for Science and Technology, Shanghai 200093, China\\
2. Department of Mathematics,  Shanghai University, Shanghai 200444, China\\ 
3. School of Mathematics, University of Leeds, Leeds LS2 9JT, UK 
}
\end{center}

{\bf Abstract:} A direct linearisation scheme, based on an elliptic Cauchy kernel, is set up for the 
lattice CKP equation. This leads to an elliptic parametrisation of the lattice CKP equation, together 
with its Lax triplet, which allows us to perform appropriate continuum limits and construct elliptic 
solutions. By selecting 
appropriate integration measures and domains for the singular linear integral equation in the scheme, 
elliptic multi-soliton solutions of the lattice CKP equation are found. 

\vspace{0.5cm} \noindent {\bf Keywords}\quad  lattice CKP equation, $\tau$-function, Lax triplet, continuum limit, elliptic soliton solution


\section{Introduction}\label{sec-1} 

The lattice CKP equation, which is connected to the hierarchy of $\mathbf{C}_\infty$ soliton 
equations of Kadomtsev-Petviashvili (KP) type, introduced in \cite{1980-JM}, was first given 
by Kashaev in \cite{1996-Kashaev-CKP}, and independently by Schief in \cite{2003-Schief-CKP}. 
The discovery of this fully discrete equation came late in comparison to other integrable 
lattice systems, because 
unlike equations of Hirota form, the $\tau$-function representation is not bilinear but a quartic 
correspondence. In the notation of the current paper the lattice CKP equation can be written as
\begin{eqnarray}  \label{eq:CKP}
\Big( {F} \wh{ \wt{\wb F}}+{\wt F} \wh{\wb F} - \wh{F} \wt{\wb F}-\wb{ F} \wh{\wt F} \Big)^2 -4\left( \wh{ \wt F} \wt{\wb F}  - { \wt F} \wh{\wt{\wb F}}  \right)  \left(\wh{ F} \wb{ F}  - { F} \wh{\wb F} \right) = 0\  ,
\end{eqnarray}
in which $F=F_{n,m,l}$ is a function of the dynamical variables (the discrete independent variables)
$n,m,l \in \mathbb Z$ and tilde-hat-bar notations
\begin{align} \label{shift-thb}
\wt{F}=F_{n+1,m,l}\ , \quad \wh{F}=F_{n,m+1,l}\ , \quad \wb{F}=F_{n,m,l+1} 
\end{align}
are used to express shifts on the $F$-function and similar undershifts such as $\dt {F}=F_{n-1,m,l}$ 
(cf.\cite{book-HJN-2016}).
The form of the equation takes the shape of a 2$\times$2$\times$2 
hyperdeterminant, due to Cayley, cf. \cite{Cayley}, and also \cite{1994-book,2007-Holtz-CKP}, if the 
lattice shifts on the 3-dimensional grid are denoted by matrix indices (like $\wt{F}=:F_{1,0,0}$, 
$\wh{F}=:F_{0,1,0}$, $\wb{F}=:F_{0,0,1}$, etc.). This intriguing form 
has led to speculations that other 
hyper-determinants may also be reinterpreted as integrable lattice equations, cf. \cite{2009-Tsarev}, but the answer to this conjecture has remained so far negative or at least inconclusive.

It is worth noting that the lattice CKP equation possesses a considerable number of compelling features. It is a direct consequence of the star-triangle relation found in connection with the 
local Yang-Baxter equation, \cite{1989-MailNij-localYB}, as was discovered in \cite{1996-Kashaev-CKP}. The left side of the equation \eqref{eq:CKP} is, in fact, the Cayley hyperdeterminant of a 2$\times$2$\times$2 cubic matrix, \cite{1994-book}. Additionally,
the  lattice CKP equation is a good example to show the connections between projective and conformal geometry and discrete integrable systems \cite{2003-Schief-CKP,2006-Schief-CKP}, and can be regarded as a special case of the hexahedron recurrence which exhibits an cluster algebra structure and  the Laurent phenomenon  \cite{2016-Kenyon-CKP}. Moreover, we note that it is related to many other  integrable systems, such as C-Toda lattice equation \cite{2018-Chang} and M-system \cite{2015-Bobenko-CKP,2017-Bobenko-CKP}. The integrablity in the sense of multi-dimensional consistency \cite{2001-Nijhoff-CAC,2002-Bobenko-MDC,2010-Doliwa-CKP} of the lattice CKP equation is concerned with the help of its relation to the symmetric principal minor assignment problem \cite{2007-Holtz-CKP,2009-Tsarev}.
The CKP equation also miraculously appears in connection 
with bi-orthogonal polynomials with Cauchy weights, \cite{2020-Chang,2025-Chang}, and the associated C-Toda systems which are connected to Degasperis-Procesi peakon dynamics, cf. \cite{2018-Chang}. 

For the sake of constructing solutions, the parameter-free version of the lattice CKP equation 
\eqref{eq:CKP} is not so useful, and in \cite{2017-DL-FN-dKP} a version was given with lattice 
parameters, i.e., parameters associated with the grid spacings, appearing in rational form. This 
explicit parametrisation facilitated the construction of in particular multi-soliton solutions. 
Furthermore, the presence of lattice parameters in the equation   
allows one also to derive in a systematic way semi-discrete and continuum analogues of the CKP 
equation, by performing well-defined limits on 
the lattice parameters and corresponding lattice variables. 

In the present paper 
we construct elliptic soliton solutions for the CKP lattice equation, and once again this is best 
done by introducing a specific elliptic parametrisation in terms of the Weierstrass $\sg$-function  
(cf. \cite{Akh} for the definition and properties), namely 
\begin{eqnarray} \label{eq:CKP-tau-f}
&&\Big( \sigma^2(\ven-\dd)\sigma^2(\nu-\ven)\sigma^2(\dd-\nu) {f} \wh{ \wt{\wb f}} +\sigma^2(\ven+\dd)\sigma^2(\nu-\ven)\sigma^2(\dd+\nu){\wt f} \wh{\wb f}  \nonumber \\  
 &&~~ - \sigma^2(\ven+\dd)\sigma^2(\ven+\nu)\sigma^2(\dd-\nu) \wh{f}\wt{\wb f} -\sigma^2(\ven-\dd)\sigma^2(\ven+\nu)\sigma^2(\dd+\nu)\wb{ f} \wh{\wt f}  \Big)^2 \nonumber \\ 
 && 
\quad=4 \sigma^2(\ven-\dd)\sigma^2(\dd+\ven)\sigma^2(\dd-\nu) \sigma^2(\dd+\nu)\left( \sigma^2(\ven+\nu) \wt{\wb f} \wh{ \wt f} -\sigma^2(\nu-\ven)\wh{\wt{\wb f}} { \wt f} \right) \nonumber \\ 
 && \quad\quad \times \left( \sigma^2(\ven+\nu) \wb{ f} \wh{ f} -\sigma^2(\nu-\ven)\wh{\wb f} {  f} \right) \  ,
\end{eqnarray}
where now $f$ denotes the dependent variable, depending like the $F$ of \eqref{eq:CKP} on lattice 
variables $n,m,l$, and   
where $\delta$, $\ven$ and $\nu$ are in the arguments of the $\sg$-functions and are the 
lattice parameters. The multidimensional consistency of this equation, embedding the equation 
in a hypercube (a tessaract) with additional lattice directions, each associated with their 
additional lattice parameter, follows from the construction of the solutions given below.

The significance of elliptic solutions to KP type and of other discrete integrable equations, not 
only lies in the fact that they form a richer class of solutions that usual multi-solitons 
(as the elliptic functions degenerate to trigonometric and rational functions under limits on the 
moduli), but in a sense the integrable equations serve as models for specific characteristics, 
such as basic and higher addition formulae for those elliptic functions, \cite{2010-Nijhoff-ellABS,2022-JNS-ellKdV}, 
as well as led to the introduction to new concepts within elliptic function theory, such as 
elliptic $N$th root of unity, \cite{2019-ellBSQ}, and bi-elliptic addition formulae, 
\cite{Atkinson-Nij2010}.

The organisation of the paper is as follows. In Section 2, we set up the direct linearisation (DL) 
scheme, involving an elliptic Cauchy kernel, that is applicable to both the AKP and CKP system. We 
introduce the main dynamical quantities and highlight which choices of so-called plane-wave 
factors apply in either case. In Section 3 we focus on the CKP case, and derive the fundamental 
relations that form the basis for the subsequent derivation of the dynamical relations for the 
main quantities. These then lead to develop the DL scheme for elliptic type solutions of the lattice CKP system and derive the Lax pair. Section 4  is dedicated to the presentation of compact and explicit forms of elliptic multi-soliton solutions. In Section 5 we present two types 
of differential-difference versions of the CKP system obtained by continuum limits: 
the straight-  and skew continuum limits leading to D$\Delta$Es with one continuum variables and two discrete variables. Finally, in the conclusion Section 6, we mention some 
further ramifications of the results in this paper.

\section{Elliptic DL scheme}\label{sec-2}
\setcounter{equation}{0}

Here we first set up the direct linearisation scheme for elliptic solutions of KP type lattice equations which holds for both AKP and CKP systems. Then in the next section we specify to the CKP 
case, as the AKP case was already covered in \cite{1983-DL-NQC}, cf. also \cite{Y-KN}. 

\subsection{Linear integral equations}\label{sec-2-1}  

The elliptic DL scheme for  the lattice CKP system commences with a pair of linear integral equations pertaining to an infinite vector denoted by  vector $\buk$  as well as its adjoint vector $\tvk$. These equations are exhibited by 
 \bse
 \label{eq:inteq}
 \be
\buk + \rho_\kp \ddint \ssl \bul \tclp {\bOm_\xi} {\boldsymbol c}_\kp = \rho_\kp \Phi_\xi(\Ld) {\boldsymbol c}_\kp\ ,
\label{eq:inteqa}
 \ee
\be
\tvk + \ssk \ddint \rho_\ell
\tvl \tcl {\bOm_\xi} {\boldsymbol c}_{\kp ^\prime}= \ssk\tck \Phi_\xi(\tLd) \ ,
\label{eq:inteqb}
\ee
\ese
in which the measure ${\rm d}\mu(\ell,\ell^\prime)$ is a function of both spectral variables $\ell$ and $\ell^\prime$, and the integration domain $D\subset \mathbb{C}^2$ is
any suitably chosen subset in the space of  the spectral variables $\ell$ and
$\ell^\prime$. 
The measure and integration domain are not explicitly specified at the beginning, but must be chosen  in a manner that ensures the uniqueness of the solution to equation \eqref{eq:inteq}. The discrete plane wave factors $ {\rho}_{\kp}={\rho}_{\kp;n,m,l}$ and $\sg_{\kp^\prime}=\sg_{\kp^\prime;n,m,l}$ are employed to demonstrate that the entries of $\buk$ and $\tvk$ are wave functions of the spectral variable and the dynamical variables $n,m,l$. The infinite column vector
$\bck=(\kp^i)_{i\in\mathbb Z}$ and the row vector $\tck=(\kp'^j)_{j\in\mathbb Z}$ are defined by listing different powers of the spectral
 variables $\kp$ and $\kp'$ respectively. Here $\Phi$ is the Lam\'e type function
\be\label{eq:Phi}
\Phi_x(y)= \frac{\sg(x+y)}{\sg(x)\sg(y)}\   . 
\ee
 The infinite matrix $\bOm_\xi$  associated with the variable $\xi$ can be viewed as a formal elliptic Cauchy kernel. Notably, unlike the rational case \cite{2017-DL-FN-dKP}, the operators $\Ld$ and $\tLd$ now appear in the linear integral equations, where they act from the left and right on $\bck$ and $\tck$, respectively, as index-raising operators, i.e.,
\be \label{c:operator}
 (\Ld\bc_\kp)_i=(\bc_\kp)_{i+1}=\kp^{i+1}\  , \quad (\tck\tLd)_{i'}=(\tck)_{i'+1}=(\kp')^{i'+1}\  .
\ee
$\Phi_\xi(\Ld), \Phi_\xi(\tLd)$ are commonly referred to as elliptic matrices which are formally coincides with substitutions of the index-raising operators in
the arguments of the relevant elliptic functions. Additionally, they share similarities with the notions of $\wp(\Ld)$, $\wp(\tLd)$, $\zeta(\Ld)$, $\zeta(\tLd)$, etc.,
where $\wp(x)$ and $\zeta(x)$ are the well-known Weierstrass functions. Finally, the formal 
\textit{elliptic Cauchy kernel} is defined as 
\be\label{eq:Cauchy}
\bOm_\xi=\colon \Phi_\xi(\Ld+\tLd))\bO\colon \ , 
\ee 
in which $\bO=\bee\tee$ (with $\bee$ an infinite vector s.t. $\tck\bee=1$, and $\tee$ a dual unit vector with $\tee\,\bck=1$) is a rank 1 projection matrix, and the \textit{normal ordering} $\colon\cdot\colon$ 
denotes that in a power series expansion of the elliptic function $\Phi_\xi$ w.r.t its argument, the 
powers of $\Ld$ appear on the right-hand side of $\bO$ while the powers of $\tLd$ appear on the left-hand 
side. 

In fact, \eqref{eq:inteq} can be simplified to a compact form by introducing the following infinite  matrices:
\be\label{eq:U-in}
{\boldsymbol U}_\xi =\ \ddint\,{\boldsymbol
u}_{\ell}\,^{t\!}{\boldsymbol c}_{\ell^\prime}\ssl \Phi_\xi(\tLd)\  ,
\ee
and 
\be\label{eq:C}
{\boldsymbol C}=\ddint\,\rho_\ell\ssl {\boldsymbol
c}_{\ell} \tclp\  .
\ee
From \eqref{eq:inteq} and  the above defination of ${\boldsymbol U}_\xi $, we have compact form for $\buk$ and $\tvk$
\bse\label{eq:uk} \bea
\buk &=& \left( \Phi_\xi(\Ld) - \bU_\xi \Phi_\xi^{-1}(\tLd) \bOm_\xi\right)
{\boldsymbol c}_\kappa \rho_\kp\ ,  \label{eq:buk} \\
\tvk&=& \sigma_{\kp'}\tck\left( \Phi_\xi(\tLd) -
\bOm_\xi \Phi_\xi^{-1}(\Ld) \bU_\xi \right)\  . \label{eq:tuk}
\eea\ese
Furthermore,  combining these formulae results in a concise expression for the infinite matrix $\bU_\xi $:
\be \label{eq:U}
\bU_\xi =  \Phi_\xi(\Ld)\, {\boldsymbol C}\, ({\boldsymbol 1}+  \bOm_\xi {\boldsymbol C})^{-1}  \Phi_\xi(\tLd)\ ,
\ee
in which $\boldsymbol 1$ is the infinite unit matrix.

In order to lay the groundwork of the elliptic DL scheme, the following aims to present the dynamical characteristics of the quantities that occur in the linear integral equations.

\subsection{ Dynamical quantities}\label{sec-2-2}

The $\rho_\kp$ and $\ssk$ are defined through the actions of 
elementary shift operators $\mathtt{T}_{\mathsf{\dd}} $, $\mathtt{T}_{\mathsf{\ven}} $, $\mathtt{T}_{\mathsf{\nu}}$ acting as follows on the plane wave factors: 
\be\label{eq:PWFshifts}
\mathtt{T}_{\mathsf{p}}\rho_\kp=\mathcal{L}_{\mathsf{p}}(\kp)\,\rho_\kp\ , \quad 
\mathtt{T}_{\mathsf{p}}\sg_{\kp^\prime}=\mathcal{R}_{\mathsf{p}}(\kp^\prime)\,
\sg_{\kp^\prime}\  , \quad {\rm with}\quad \mathsf{p}=\delta, \ven, \nu\ ,   
\ee 
in which the factors 
$\mathcal{L}_{\mathsf{p}}(\kp)$ and $\mathcal{R}_{\mathsf{p}}(\kp^\prime)$ will be specified below. 
In what follows we shall use the following abbreviations:  
\bse\label{eq:rho}\bea \label{eq:rhoa}
& \wt{\rho}_\kp=\mathtt{T}_{\mathsf{\dd}} \rho_\kp \ ,  \quad \wh{\rho}_\kp=\mathtt{T}_{\mathsf{\ven}} \rho_\kp\    \ ,  \quad    \wb{\rho}_\kp=\mathtt{T}_{\mathsf{\nu}} \rho_\kp\    \ , \\ 
& \wt{\sg}_{\kp^\prime}=\mathtt{T}_{\mathsf{\dd}} \sg_{\kp^\prime} \   , \quad 
\wh{\sg}_{\kp^\prime}=\mathtt{T}_{\mathsf{\ven}} \sg_{\kp^\prime} \ ,\quad 
\wb{\sg}_{\kp^\prime}=\mathtt{T}_{\mathsf{\nu}} \sg_{\kp^\prime}\   ,
\eea\ese 
in which $\mathtt{T}_{\mathsf{\dd}} $ is a forward shift operation characterised by the lattice parameter $\dd$. We will also occasionally employ the notation of under-accents to denote backward shifts, i.e.
\[  {\hypotilde 0 \rho}_\kp=\mathtt{T}^{-1}_{\mathsf{\dd}} \rho_\kp \ , \quad {\hypohat 0 \rho}_\kp=\mathtt{T}^{-1}_{\mathsf{\ven}} \rho_\kp \ , \quad 
{\hypobar 0 \rho}_\kp=\mathtt{T}^{-1}_{\mathsf{\nu}} \rho_\kp \ ,
\]
and similarly for the actions of inverse shifts on $\sg_{\kp^\prime}$, and on all other 
derived quantities.

The above shift relations under translations along the lattice and index-raising operators implies the relations for $\bc_\kp$ and $\tck$, 
\be\label{eq:ck-shift}
(\mathtt{T}_{\mathsf{p}} {\rho}_\kp)\bc_\kp = {\rho}_\kp
\mathcal{L}_{\mathsf{p}}(\Ld)\,\bc_\kp\   , \quad  
\tck (\mathtt{T}_{\mathsf{p}}^{-1}\ssk)  =\tck  
\mathcal{R}_{\mathsf{p}}(\tLd)\,\ssk \   , 
\quad \mathsf{p}=\dd,\ven,\nu \  . 
\ee
Eq. \eqref{eq:ck-shift} implies the following shift relations for the 
matrix  ${\boldsymbol C}$ of \eqref{eq:C}, 
\be\label{eq:C-shift}
\mathtt{T}_{\mathsf{p}} ({\boldsymbol C}) \cdot \mathcal{R}_{\mathsf{p}} = \mathcal{L}_{\mathsf{p}}  \cdot  {\boldsymbol C}\   .  
\ee
Furthermore, the key property of the formal elliptic Cauchy kernel $\bOm_\xi$ is the (infinite) 
compatible set of relations:
\be\label{eq:Omega}
\mathtt{T}_{\mathsf{p}} ({\bOm_\xi})\mathcal{L}_{\mathsf{p}}(\Ld) -\mathcal{R}_{\mathsf{p}}(\tLd) \bOm_{\xi} =
\mathcal{O}_{\mathsf{p}} \   ,
\ee
in which $\mathcal{O}_{\mathsf{p}}$ are finite-rank projector labelled by $\mathsf{p}$.  
Henceforth we will simply write $\mathcal{R}_{\mathsf{p}}$ and $\mathcal{L}_{\mathsf{p}}$
for the operators $\mathcal{R}_{\mathsf{p}}(\tLd)$ and $\mathcal{L}_{\mathsf{p}}(\Ld)$ 
respectively. Eq. \eqref{eq:Omega} could be thought of as a definition of $\bOm_\xi$ 
(instead of \eqref{eq:Cauchy}), 
noting that it is nontrivial to find operators $\mathcal{R}_{\mathsf{p}}$ and $\mathcal{L}_{\mathsf{p}}$, and corresponding projector $\mathcal{O}_{\mathsf{p}}$, s.t. 
the relation holds for all $\mathsf{p}$ simultaneously, while $\bOm_\xi$ does not 
depend explicitly on $\mathsf{p}$. If we find such a solution, the construction that follows 
guarantees the multidimensional consistency of the corresponding equations that are 
covariantly constructed.

By utilizing the formulae \eqref{eq:C-shift} and \eqref{eq:Omega},  it is possible to derive dynamic relations for the infinite matrix $\bU_\xi$ from \eqref{eq:U}. These relations will prove useful in the construction of the lattice CKP equation. To do that, we also need to introduce certain scalar functions:
\begin{itemize}  
\item
the $\tau$-function given by
the infinite determinant
\be\label{eq:tau}
\tau = \det\,_{{\mathbb Z}\times {\mathbb Z}}\left(
{\boldsymbol 1}+\bOm_\xi\cdot\bC\right)\  .
\ee
The determinant could be understood as the expansion of ${\rm exp}\{{\rm tr}[{\rm ln}({\boldsymbol 1}+\bOm_\xi\cdot\bC)]\}$.
\item a central function $u=\tee\,\bU_\xi\bee$\ .
\item  $\ar,\beta$ parameter-dependent quantities
\bse\label{eq:vws}\bea
v_\ar &=& 1 - \tee\,\bh^{-1}( \xi,\ar,\Ld) \cdot \bU_\xi\,\bee 
\   ,  \label{eq:v} \\
w_\ar &=& 1 - \tee\,\bU_\xi\cdot \boldsymbol h^{-1}( \xi,\ar,\tLd)\,\bee
\   ,  \label{eq:w}\\
s_{\ar,\bb} &=& \tee\,\bh^{-1}( \xi,\ar,\Ld)\cdot\bU_\xi \cdot \boldsymbol h^{-1}( \xi,\beta,\tLd)\,\bee \   ,  \label{eq:s}
\eea \ese
with \be \boldsymbol h(\alpha,\beta,x)= \zeta(\alpha)+\zeta(\beta)+\zeta(x)-\zeta(\alpha+\beta+x)
=\frac{\Phi_{\alpha}(x)\,\Phi_\beta(x)}{\Phi_{\alpha+\beta}(x)}\  , \label{chi-1}
\ee
\end{itemize}
where $\tee$ and $\bee$ are the infinite vectors associated with the rank 1 projector $\bO$. 
The operators involving the $\bh$ function are interpreted as formal matrices 
$\boldsymbol h( \xi,\ar,\Ld)= \zeta(\xi) {\boldsymbol 1} +\zeta(\ar) {\boldsymbol 1} +\zeta( \Ld)-\zeta(\xi{\boldsymbol 1}+\ar{\boldsymbol 1}+ \Ld), $ etc., involving analytic functions of the 
shift matrices $\Ld$ and $\tLd$. 
 In what follows we will, for notational convenience omit the infinite unit matrix symbol, 
 simply only write the scalar coefficient where it is understood that they are 
 multiplied by ${\boldsymbol 1}$. 

 For computational purposes, the function $\boldsymbol h$ obeys the identity:
\be\label{eq:h-identity}
\frac{\boldsymbol h(\xi,\delta,\lambda)}{\boldsymbol h(\xi,\alpha,\lambda)} 
=1+ \frac{\zeta(\delta)+\zeta(\xi+\alpha)-\zeta(\alpha)-\zeta(\xi+\delta)}{\boldsymbol h(\xi+\delta,\alpha,\lambda)}\ ,
\ee 
which was repeatedly used in \cite{2019-ellBSQ}, cf. also \cite{Y-KN}, in the derivation of the key relations underlying the AKP structure. For the CKP system of equations this identity is supplemented with \eqref{eq:hh-identity} below relevant for the derivations in the CKP case.  
In the following,  we will conduct a detailed analysis of the elliptic DL scheme associated with the CKP lattice equation. This will be achieved by providing an explicit representation of equation \eqref{eq:Omega}.

\section{The lattice CKP equation} 
\setcounter{equation}{0}

The DL scheme presented in the previous section applies to both the AKP system as well as to the CKP system, by appropriate choices of the operators $\mathcal{R}_{\mathsf{p}}$ and $\mathcal{L}_{\mathsf{p}}$. In fact, for the AKP, and CKP systems we have the following choices\footnote{The BKP system has a different Cauchy kernel from the AKP and CKP system, cf. 
\cite{2017-DL-FN-dKP}, hence we don't include the BKP system here.}: 
\begin{itemize}
\item{AKP system:} 
\be
\mathcal{L}_{\mathsf{p}}=\Phi_{\mathsf{p}}(\Ld)\ , \quad \mathcal{R}{\mathsf{p}}=\Phi_{\mathsf{p}}(-\tLd)\ , \quad \quad \mathsf{p}=\dd,\ven,\nu \ , \ee 
while $\mathcal{O}_{\mathsf{p}} =\bO$, and $\xi=\xi_0-n\delta-m\varepsilon-l\nu$, where $\xi_0$ is fixed. There is no 
restriction on the measure ${\rm d}\mu(\ell,\ell^\prime)$. 
\item{CKP system:}
\be
\mathcal{L}_{\mathsf{p}}=\frac{\Phi_{\mathsf{p}}(\Ld)}{\Phi_{\mathsf{p}}(-\Ld)}, \quad \mathcal{R}{\mathsf{p}}=\frac{\Phi_{\mathsf{p}}(-\tLd)}{\Phi_{\mathsf{p}}(\tLd)}\ , \, \ee  
while 
\be \mathcal{O}_{\mathsf{p}} =\sg(2\mathsf{p})\Phi_{\mathtt{T}_{\mathsf{p}}\xi}(\tLd+\mathsf{p})\bO\Phi_{\xi}(\Ld-\mathsf{p})\ ,\quad \mathsf{p}=\dd,\ven,\nu \ , 
\ee\ and $\xi=\xi_0-2n\delta-2m\varepsilon-2l\nu$. The measure (and integration domain) are 
symmetric, i.e. ${\rm d}\mu(\ell,\ell^\prime)={\rm d}\mu(\ell^\prime,\ell)$. 
\end{itemize}

In the CKP case, the symmetry condition imposed on the integration measure 
and integration domain of the spectral parameters $\ell,\ell^\prime$ together 
with the transposition symmetry on the formal Cauchy kernel $\bOm_{\xi}$, 
implies the symmetry 
\[ \bC=\,^t\!\bC  \ ,\quad \bU_\xi=\,^t\!\bU_\xi \]
on the infinite matrix $\bU_\xi$. This in turn implies that 
the quantities defined in \eqref{eq:vws} inherit this symmetry, i.e., 
\be
{v}_\ar= w_{\ar}\, \quad \text{and}    \quad s_{\ar,\bb} =s_{\bb,\ar} \  .
\ee
This will be taken into account in the ensuing analysis.

\subsection{Basic shift relations for the parameter-dependent quantities - CKP case} 

Focusing now exclusively on the CKP case we have the following central relation for 
the elliptic Cauchy kernel: 
\be\label{eq:Omega-CKP-dd}
\wt{\bOm_\xi}\,\frac{\Phi_{\dd}(\Ld)}{\Phi_{\dd}(-\Ld)} - 
\frac{\Phi_{\dd}(-\tLd)}{\Phi_{\dd}(\tLd)}\,\bOm_{\xi} =
\sg(2\dd)\,\Phi_{\wt\xi}(\tLd+\dd)\,\bO\,\Phi_{\xi}(\Ld-\dd)  \   ,
\ee 
which, together with \eqref{eq:Cauchy} follows from the  specific elliptic identity:  
\begin{align}\label{eq:ellid1}
& \Phi_\delta(\lambda)\,\Phi_\delta(\kappa)\,\Phi_{\xi-2\delta}(\kappa+\lambda) 
- \Phi_\delta(-\lambda)\,\Phi_\delta(-\kappa)\,\Phi_{\xi}(\kappa+\lambda)  \\ 
&= \Phi_{\xi-\delta}(\lambda)\,\Phi_{\xi-\delta}(\kappa)\,\frac{\sigma(2\delta)\,\sigma^2(\xi-\delta)}{\sigma(\xi)\,\sigma^2(\delta)\,\sigma(\xi-2\delta)}
= -\sigma(2\delta)\,\Phi_{-\delta}(\kappa)\,\Phi_\delta(\lambda)\,\Phi_{\xi-2\delta}(\lambda+\delta)
\,\Phi_\xi(\kappa-\delta)\ . \nonumber 
\end{align}
The dynamics in the $\mathsf{p}=\delta$ direction is given by 
\be\label{eq:CCa-CKP}
\wt{\boldsymbol C}\frac{\Phi_\dd(-\tLd)}{\Phi_\dd(\tLd)}=\frac{\Phi_\dd(\Ld)}{\Phi_\dd(-\Ld)} {\boldsymbol C}  \   ,   
\ee 
while 
\be\label{eq:xi}
\xi=\xi_0-2n\delta-2m\varepsilon-2l\nu \ \quad \Rightarrow\quad \wt{\xi}=\xi-2\delta\ . 
\ee

Given these relations, we will see how the elliptic DL scheme is able to both formulate and resolve the lattice CKP equation. Our initial step involves identifying the dynamical relationships inherent in the $\tau$-function. By implementing a~~$\wt{}$~~shift on \eqref{eq:tau} and considering the equations \eqref{eq:CCa-CKP} and \eqref{eq:Omega-CKP-dd}, we thus derive the following:
\be\label{shift-tau-CKP}   
\wt{\tau}
= \left(1- \boldsymbol h( \xi,-\dd,-\dd) s_{-\dd,-\dd}\right) {\tau}  \   ,
\ee
and apply the same methodology to deduce relations for the remaining lattice directions.  On the other hand, we obtain
 \be\label{shift-tau-CKP-dt}
 \ut{\tau}=\left(1- \boldsymbol h( \xi,\dd,\dd\right)s_{\dd,\dd}){\tau} \   ,
\ee
and establish analogous shift relations for other lattice directions.

Next we discuss dynamical relations of the quantities ${v}_\ar$ and $s_{\ar,\beta}$ defined in 
\eqref{eq:vws}. First, we establish the key dynamical relation for the infinite matrix $\bU_\xi$ 
which is readily derived as 
\be \label{eq:Urel-CKP} 
\wt{\bU_\xi} \frac{\boldsymbol h(\xi,-\delta,\tLd)}{\boldsymbol h(\wt\xi,\delta,\tLd)} 
= \frac{\boldsymbol h(\wt\xi,\delta,\Ld)}{\boldsymbol h(\xi,-\delta,\Ld)}\bU_\xi -
\boldsymbol h(\wt\xi,\dd,\dd)\, \wt{\bU_\xi} \frac{1}{\boldsymbol h(\wt\xi,\delta,\tLd)} 
\bO \frac{1}{\boldsymbol h(\xi,-\delta,\Ld)}\bU_\xi  \   , 
 \ee
 and similar relations for the shifts of $\bU_\xi$ in the other lattice directions. Eq. 
\eqref{eq:Urel-CKP} forms the starting point for the subsequent analysis and construction of 
nonlinear relations for the $v_\ar$ and $s_{\ar,\beta}$ by operating on the left right by operators of the type $\bh(\wt{\xi},\ar,\Ld)^{-1}$ and/or  $\bh(\xi,\beta,\tLd)^{-1}$, 
and sandwiching the results between the vectors $\tee$ and $\bee$, making use of the 
symmetry $\,^t\!\bU_\xi=\bU_\xi$ induced by the symmetry of the measure and integration domain.  

From \eqref{eq:Urel-CKP}, sandwiching the relation between the vectors $\tee$ and $\bee$, we obtain the following relation
\be\label{eq:uvrel}
\bh(\wt{\xi},\dd,\dd)+u-\wt{u}=\bh(\wt{\xi},\dd,\dd)\,\wt{v_\dd}\,v_{-\dd}\ . 
\ee 
Left-multiplying $\boldsymbol h^{-1}(\wt \xi,\ar,\Ld)$ to \eqref{eq:Urel-CKP} and taking the central entry yields
\begin{eqnarray}
 \wt{v_{\al}}-\frac{\boldsymbol h(\xi,-\dd,\al)}{\boldsymbol h(\wt{\xi},\dd,\al)}v_\al=\boldsymbol h(\xi,-\dd,-\dd)\Big( \wt{s_{\al,\dd}}-\frac{1}{\boldsymbol h(\wt{\xi},\dd,\al)} \Big) v_{-\dd}~, \label{eq:ua-dyna-a}
\end{eqnarray}
where use has been made of the following key identity, which is proven in the Appendix, 
\be\label{eq:hh-identity}
\frac{\boldsymbol h(\wt{\xi},\delta,\lambda)}{\boldsymbol h(\wt{\xi},\alpha,\lambda)\,\boldsymbol h(\xi,-\delta,\lambda)}=
\frac{1}{\bh(\wt{\xi},\delta,\ar)}
\left(\frac{\bh(\xi,\ar,-\delta)}{\bh(\xi,\ar,\ld)}-\frac{\bh(\xi,-\delta,-\delta)}{\bh(\xi,-\delta,\ld)}\right)\ , 
\ee 
and which constitutes an elliptic partial fraction expansion in terms of the 
function $\bh$ and w.r.t. the variable $\ld$ (which represents the 
matrix $\Ld$ or $\tLd$ in the derivations).

Setting $\al=\dd$ in \eqref{eq:ua-dyna-a} we find
\begin{eqnarray}
\frac{\wt{v_{\dd}}}{{v}_{-\dd}} =1-\boldsymbol h(\wt \xi,\dd,\dd) \wt{s_{\dd,\dd}}=
\frac{\tau}{\wt{\tau}}\  ,  \label{eq:V-dyna-a}
\end{eqnarray}
where the latter equality follows from \eqref{shift-tau-CKP}. 

A second type of relations is obtained from multiplying both sides of the relation for $\bU_\xi$, namely considering the central element of $\boldsymbol h^{-1}(\wt \xi,\ar,\Ld) \cdot\eqref{eq:Urel-CKP} \cdot \boldsymbol h^{-1}( \xi,\beta,\tLd)$, which leads to a relation purely
in terms of the objects ${s}_{\al,\bb}$
\begin{eqnarray}\label{eq:S-dyna-abp}
\frac{1}{\boldsymbol h(\xi,-\dd,-\dd)} \big(\boldsymbol h(\xi,\al,-\dd)\boldsymbol h(\xi,-\dd,\bb){s}_{\al,\bb}- \boldsymbol h(\wt \xi,\al,\dd) \boldsymbol h(\wt \xi,\dd,\bb) \wt{s_{\al,\bb}}   \big)  \nn\\
\quad+ \big(1-\boldsymbol h(\wt \xi,\al,\dd) \wt{s_{\al,\dd}} \big)\big(1- \boldsymbol h(\xi,-\dd,\bb)s_{-\dd,\bb} \big)= 1 \  , 
\end{eqnarray}
in which use has been made of \eqref{eq:hh-identity} in combination with a similar identity involving 
a parameter $\bb$ instead of $\ar$, namely 
\be\label{eq:hhh-identity}
\frac{\bh(\xi,-\delta,\lambda)}{\bh(\xi,\beta,\lambda)\,\boldsymbol h(\wt{\xi},\delta,\lambda)}=
\frac{1}{\bh(\xi,-\delta,\beta)}
\left(\frac{\bh(\wt{\xi},\beta,\delta)}{\bh(\wt{\xi},\beta,\ld)}-\frac{\bh(\wt{\xi},\delta,\delta)}{\bh(\wt{\xi},\delta,\ld)}\right)\ , 
\ee 
which is obtained from the former by changing $\delta\to -\delta$, and by 
reversing the lattice shifts.  
The relations \eqref{eq:ua-dyna-a} and \eqref{eq:S-dyna-abp} together, for different choices of the parameters 
$\ar$ and $\beta$ plus their analogies in the other two lattice directions, are enough to derive 
the lattice CKP equation. First, we enumerate several special cases by choosing particular parameters in the relation \eqref{eq:S-dyna-abp}, namely: 
\begin{itemize}
\item $\ar=\dd$
\begin{align}\label{eq:S-dyn-bp}
\boldsymbol h(\wt \xi,\dd,\bb) \wt{s_{\dd,\bb} }+\big(1-\boldsymbol h(\wt \xi,\dd,\dd) \wt{s_{\dd,\dd}} \big)\big(1- \boldsymbol h(\xi,-\dd,\bb)s_{-\dd,\bb} \big)= 1 \  . 
\end{align}

\item $\ar=\dd, \, \beta=-\dd$  
\begin{align} 
\left(1- \boldsymbol h( \xi,-\dd,-\dd) s_{-\dd,-\dd}\right) \left(1- \boldsymbol h( \wt \xi,\dd,\dd)\wt{s_{\dd,\dd}} \right)=1 \  .
\end{align}
This can be seen from combining \eqref{shift-tau-CKP} and \eqref{shift-tau-CKP-dt}.
\item  $\ar=\dd, \, \beta=\ven$  
\begin{eqnarray} \label{eq:S-dyna-dv-1}
 \frac{1-\boldsymbol h(\wt \xi,\dd,\ven) \wt{s_{\dd,\ven}}}{1- \boldsymbol h(\xi,-\dd,\ven)s_{-\dd,\ven} }= \frac{\tau}{\wt{\tau}} \  .
\end{eqnarray}

\item  $\ar=\dd, \, \beta=\nu$  
 \begin{eqnarray}  \label{eq:S-dyna-dn-1}
 \frac{1-\boldsymbol h(\wt \xi,\dd,\nu) \wt{s_{\dd,\nu}}}{1- \boldsymbol h(\xi,-\dd,\nu)s_{-\dd,v} }= \frac{\tau}{\wt{\tau}} \  . 
\end{eqnarray}
The above two formulae follow from \eqref{eq:S-dyna-abp} and \eqref{shift-tau-CKP-dt} $\wt{}$ \   .
\item $\beta= -\dd $
\begin{eqnarray}
\big(1-\boldsymbol h(\wt \xi,\al,\dd) \wt{s_{\al,\dd}} \big)\big(1- \boldsymbol h(\xi,-\dd,-\dd)s_{-\dd,-\dd} \big)+ \boldsymbol h(\xi,\al,-\dd){s}_{\al,-\dd}  = 1\  . \label{eq:S-dyna-ap-}
\end{eqnarray}
This reduces to
\begin{eqnarray}
1- \boldsymbol h(\xi,-\dd,-\dd)s_{-\dd,-\dd} =\frac{1- \boldsymbol h(\xi,\al,-\dd){s}_{\al,-\dd}}{1-\boldsymbol h(\wt \xi,\al,\dd) \wt{s_{\al,\dd}}}= \frac{\wt \tau}{{\tau}}\  , \label{eq:S-dyna-ap}
\end{eqnarray}
benefiting from  \eqref{shift-tau-CKP}.
\item $\al=\bb=\ven$
\begin{eqnarray}
\frac{1}{\boldsymbol h(\wt \xi,\dd,\dd)} \Big(\boldsymbol h^{2 }(\wt \xi,\dd,\ven) \wt{s_{\ven,\ven}} -\boldsymbol h^{2 }(\xi,-\dd,\ven){s}_{\ven,\ven} \Big)  + \,  \big(1-\boldsymbol h(\wt \xi,\ven,\dd) \wt{s_{\ven,\dd}} \big)\big(1- \boldsymbol h(\xi,-\dd,\ven)s_{-\dd,\ven} \big) 
 = 1 \  .  \nn
\end{eqnarray}
This turns into 
\begin{eqnarray} \label{eq:S-dyn-pq2}
\Big(1-\boldsymbol h(\wt \xi,\dd,\ven) \wt{s_{\dd,\ven}} \Big)^2  \frac{\wt \tau}{\tau}=1+  \frac{1}{\boldsymbol h(\xi,-\dd,-\dd)} \Big(\boldsymbol h^{2 }(\wt \xi,\dd,\ven) \wt{s_{\ven,\ven}} - \boldsymbol h^{2 }(\xi,-\dd,\ven){s}_{\ven,\ven} \Big) \ ,
\end{eqnarray}
taking advantage of the relation \eqref{eq:S-dyna-dv-1}  and the symmetry property of ${s}_{\al,\bb}$.

\item $\al=\ven,\bb=\nu$ 
\begin{eqnarray}
&&\big(1-\boldsymbol h(\wt \xi,\ven,\dd) \wt{s}^{}_{\ven,\dd} \big)\big(1- \boldsymbol h(\xi,-\dd,\nu)s^{}_{-\dd,\nu} \big)   \nonumber \\
&+&\frac{1}{\boldsymbol h(\xi,-\dd,-\dd)} \big( \boldsymbol h(\wt \xi,\ven,\dd) \boldsymbol h(\wt \xi,\dd,\nu) \wt{s}^{}_{\ven,\nu} - \boldsymbol h(\xi,\ven,-\dd)\boldsymbol h(\xi,-\dd,\nu){s}^{}_{\ven,\nu} \big) = 1\  .\label{eq:S-dyna-pqr-1}
\end{eqnarray}
This can be rearranged as 
\begin{eqnarray}
&&\big(1-\boldsymbol h(\wt \xi,\ven,\dd) \wt{s^{}_{\ven,\dd}} \big) \big(1-\boldsymbol h(\wt \xi,\dd,\nu) \wt{s^{}_{\dd,\nu}}\big) \frac{\wt\tau}{{\tau}}+\frac{\boldsymbol h(\wt \xi,\ven,\dd) \boldsymbol h(\wt \xi,\dd,\nu)}{\boldsymbol h(\wt \xi,\ven,\nu) \boldsymbol h(\wt \xi,\dd,\dd)} \big(  1-\boldsymbol h(\wt \xi,\ven,\nu)\wt{s^{}_{\ven,\nu}}\big) 
   \nonumber \\
&&\quad \quad -\frac{\boldsymbol h(\xi,\ven,-\dd)\boldsymbol h(\xi,-\dd,\nu)}{\boldsymbol h(\xi,\ven,\nu)\boldsymbol h(\xi,-\dd,-\dd)} \big(1-  \boldsymbol h(\xi,\ven,\nu){s}^{}_{\ven,\nu} \big) = 0\  , \quad\label{eq:S-dyna-pqr-2}
\end{eqnarray}
 by making use of \eqref{eq:S-dyna-dn-1} and the relation (which follows straightforwardly from the 3-term relation for the Weierstrass $\sg$-function) 
\be
1-\frac{\boldsymbol h(\wt \xi,\ven,\dd)\boldsymbol h(\wt \xi,\dd,\nu)}{\boldsymbol h(\wt \xi,\dd,\dd)\boldsymbol h(\wt \xi,\ven,\nu)}  - \frac{\boldsymbol h(\xi,\ven,-\dd)\boldsymbol h(\xi,-\dd,\nu)}{\boldsymbol h(\xi,-\dd,-\dd)\boldsymbol h(\xi,\ven,\nu)}=0\  . \label{re-kai-2}
\ee 
\end{itemize} 

Now we have some shift relations for $s_{\al,\bb}$ and $\tau$-function in $n$-direction. The shift relations with respect to $\{\ven,~\wh{}~\}$ or $\{\nu,~\bar{}~\}$ can be obtained by interchanging $\{\dd,~\wt{}~\}$ with $\{\ven,~\wh{}~\}$ or $\{\nu,~\bar{}~\}$ in the relations \eqref{shift-tau-CKP}, \eqref{eq:S-dyna-abp} and relations derived from them. In general, these shift relations can be applied to find particular discrete integrable systems.

\subsection{The lattice CKP equation}
In an effort to derive a closed-form equation for the $\tau$-function, we will resort to the above shift relations in the following. We first notice that ${s}_{\ven,\ven}$ in \eqref{eq:S-dyn-pq2} can be eliminated by  the counterpart of \eqref{eq:S-dyn-bp}
for the $~\dh{}~$-shift, which reads
\begin{eqnarray}
{s}^{}_{\ven,\ven} =\frac{1}{\boldsymbol h(\xi,\ven,\ven)} \Big(1- \frac{\dh \tau}{{\tau}}\Big) \  .\label{eq:S-dyna-bp}
\end{eqnarray} 
Then we have
\begin{eqnarray}
\Big(1-\boldsymbol h(\wt \xi,\dd,\ven) \wt{s^{}_{\dd,\ven}} \Big)^2  \frac{\wt \tau}{\tau}=1- \frac{\boldsymbol h^{2 }(\wt \xi,\dd,\ven)}{\boldsymbol h(\wt \xi,\dd,\dd)\boldsymbol h(\wt \xi,\ven,\ven)} \Big(1- \frac{\wt{\dh \tau}}{\wt{\tau}}\Big) - \frac{\boldsymbol h^{2 }(\xi,-\dd,\ven)}{\boldsymbol h(\xi,-\dd,-\dd)\boldsymbol h(\xi,\ven,\ven)} \Big(1- \frac{\dh \tau}{{\tau}}\Big) \   . \quad \label{eq:S-dyna-pq2}
\end{eqnarray}
Using  $\sigma$-three term  relation, we know that 
\be
1-\frac{\boldsymbol h^{2 }(\wt \xi,\dd,\ven)}{\boldsymbol h(\wt \xi,\dd,\dd)\boldsymbol h(\wt \xi,\ven,\ven)}  - \frac{\boldsymbol h^{2 }(\xi,-\dd,\ven)}{\boldsymbol h(\xi,-\dd,-\dd)\boldsymbol h(\xi,\ven,\ven)}=0\  .  \label{re-kai-1}
\ee
\eqref{eq:S-dyna-pq2} can be then rewritten as
\bse\label{rel:CKP-f}
\begin{eqnarray}\label{eq:S-dyna-dv-3}
\Big(1-\boldsymbol h(\xi,\dd,\ven) {s}^{}_{\dd,\ven} \Big)^2  &=&\sg^2(\xi) \frac{\sg^2(-\dd-\ven)\dt f \dh f-\sg^2(-\dd+\ven)\dh{\dt f}  f }{\sg^2(\xi+\dd+\ven)\sg(-2\dd)\sg(-2\ven)f^2}\  ,
\end{eqnarray}
in which $f= \sg(\xi)\tau$. Similarly, we have its counterparts in the form
\begin{eqnarray}
\Big(1-\boldsymbol h(\xi,\dd,\nu) {s}^{}_{\dd,\nu} \Big)^2  &=& \sg^2(\xi) \frac{\sg^2(\dd+\nu)\dt f \underline f-\sg^2(-\nu+\dd)\underline{\dt f}  f }{\sg^2(\xi+\dd+\nu)\sg(2\dd)\sg(2\nu)f^2}\  , \\
\Big(1-\boldsymbol h( \xi,\nu,\ven) {s}^{}_{\nu,\ven} \Big)^2  &=& \sg^2(\xi) \frac{\sg^2(\nu+\ven)\underline f \dh f-\sg^2(-\ven+\nu)\dh{\underline f}  f }{\sg^2(\xi+\nu+\ven)\sg(2\nu)\sg(2\ven)f^2}\  .
\end{eqnarray}
\ese
Variables ${s}^{}_{\dd,\ven}, {s}^{}_{\nu,\ven}, {s}^{}_{\nu,\dd}$ in the equation \eqref{eq:S-dyna-pqr-2} can be expressed in terms of $f$-function by relations \eqref{rel:CKP-f}, from which we obtain the closed quartic form 
\begin{eqnarray} \label{eq:CKP-tau-ff}
&&\Big( (\Phi_{-\dd}(\ven))^2 ( \Phi_{-\ven}(\nu))^2 ( \Phi_{-\nu}(\dd))^2 {f} \wb{ \wh{\wt f}} +( \Phi_{\dd}(\ven))^2 ( \Phi_{-\ven}(\nu))^2( \Phi_{\nu}(\dd))^2{\wt f} \wh{\wb f}  \nonumber \\  
 &&~~ - ( \Phi_{\dd}(\ven))^2( \Phi_{\nu}(\ven))^2( \Phi_{-\nu}(\dd))^2 \wh{f}\wb{\wt f} -( \Phi_{-\dd}(\ven))^2( \Phi_{\nu}(\ven))^2( \Phi_{\nu}(\dd))^2\wb{ f} \wh{\wt f}  \Big)^2 \nonumber \\ 
 && 
\quad=4( (\Phi_{-\dd}(\ven))^2( (\Phi_{\dd}(\ven))^2 ( \Phi_{-\nu}(\dd))^2 ( \Phi_{\nu}(\dd))^2\left( ( \Phi_{\nu}(\ven))^2 \wb{\wt f} \wh{ \wt f} -( \Phi_{-\ven}(\nu))^2 \wb{\wh{\wt f}} { \wt f} \right) \nonumber \\ 
 && \quad\quad \times \left( ( \Phi_{\nu}(\ven))^2 \wb{ f} \wh{ f} -( \Phi_{-\ven}(\nu))^2 \wb{\wh f} {  f} \right) \ ,
\end{eqnarray}
which simplifies to \eqref{eq:CKP-tau-f}. By means of the transformation  
\begin{eqnarray}
f= \left(\frac{\sg(-\dd-\ven)}{\sg(-\dd+\ven)}\right)^{2nm}\left(\frac{\sg(-\ven-\nu)}{\sg(-\ven+\nu)}\right)^{2ml}\left(\frac{\sg(-\nu-\dd)}{\sg(-\nu+\dd)}\right)^{2nl} F\ , 
\end{eqnarray}
we remove the lattice parameters $\dd,\ven,\nu$ in \eqref{eq:CKP-tau-f} and see that $F$ satisfies the lattice CKP equation \eqref{eq:CKP}. 

In addition to the CKP equation for the $\tau$-function $f$ we have also a coupled system of lattice 
equations for the quantities $u$, $v_\ar$, $v_\beta$, $s_{\ar,\beta}$, where we eliminated the 
dependent variables carrying the labels associated with the lattice parameters (such as $v_\delta$, $v_{-\dd}$, $s_{\ar,\dd}$, etc.), leading to 
\bse\label{eq:uvssyst}\begin{align}
  &  \boldsymbol h(\xi,\al,-\dd)\boldsymbol h(\xi,-\dd,\bb){s}_{\al,\bb}+ \boldsymbol h(\wt \xi,\al,\dd) \boldsymbol h(\wt \xi,\dd,\bb) \wt{s_{\al,\bb}}    \nonumber \\
  &  \quad -\frac{[\bh(\xi,\ar,-\dd)v_\ar-\bh(\wt{\xi},\ar,\dd)\wt{v_\ar}]\,[\bh(\xi,\beta,-\dd)v_\beta-\bh(\wt{\xi},\beta,\dd)\wt{v_\beta}]}{u-\wt{u}+\bh(\wt{\xi},\dd,\dd)}=  \boldsymbol h(\xi,-\dd,-\dd)\ , \label{eq:uvssyst-t} \\ 
   &  \boldsymbol h(\xi,\al,-\ven)\boldsymbol h(\xi,-\ven,\bb){s}_{\al,\bb}+ \boldsymbol h(\wh \xi,\al,\ven) \boldsymbol h(\wh \xi,\ven,\bb) \wh{s_{\al,\bb}}   \nonumber \\
  &  \quad-\frac{[\bh(\xi,\ar,-\ven)v_\ar-\bh(\wh{\xi},\ar,\ven)\wh{v_\ar}]\,[\bh(\xi,\beta,-\ven)v_\beta-\bh(\wh{\xi},\beta,\ven)\wh{v_\beta}]}{u-\wh{u}+\bh(\wh{\xi},\ven,\ven)}=\boldsymbol h(\xi,-\ven,-\ven)\ , \\ 
   &  \boldsymbol h(\xi,\al,-\nu)\boldsymbol h(\xi,-\nu,\bb){s}_{\al,\bb}+ \boldsymbol h(\wb\xi,\al,\nu) \boldsymbol h(\wb \xi,\nu,\bb) \wb{s_{\al,\bb}}    \nonumber \\
  & \quad -\frac{[\bh(\xi,\ar,-\nu)v_\ar-\bh(\wb{\xi},\ar,\nu)\wb{v_\ar}]\,[\bh(\xi,\beta,-\nu)v_\beta-\bh(\wb{\xi},\beta,\nu)\wb{v_\beta}]}{u-\wb{u}+\bh(\wb{\xi},\nu,\nu)}=\boldsymbol h(\xi,-\nu,-\nu)\ .   
\end{align}\ese 
It may be possible to reduce this system further, by eliminating variables using intermediate relations and combining the 
shifts, but so far we have not found any significant simplification. 
Setting $\alpha=\beta$ in \eqref{eq:uvssyst} we get a system of three coupled O$\Delta$Es for the 
three quantities $u$, $v_\ar$ and $s_{\ar,\ar}$ in the three lattice directions separately. There 
should be further relations, in fact P$\Delta$Es in two shifts, arising from the compatibility  of the
O$\Delta$Es. Once solved, by back-substitution of the solutions of $u$, $v_\ar$, and $v_\bb$ 
(obtained from the similar system replacing $\ar$ by $\beta$) we should be able to find $s_{\ar,\bb}$ 
for $\ar\neq\bb$. We leave this analysis to a future work.

\subsection{Lax triplet}

The elliptic DL can also lead to Lax pairs of  the equation that we are investigating, which appears in the form of a triplet which 
replaces the conventional Lax pairs. In the part, we will demonstrate the Lax pair of the lattice CKP equation. This, in fact,  requires the shift relation of $\buk$, which is acquired from the discrete plane wave factors and the dynamical relations of $\bOm_\xi$ and $\bU_\xi$. To be more specific, making use of the relation
\be \label{eq:rhoa-CKP}
 \rho_\kp\ \ \to\ \
\wt{\rho}_\kp=\mathtt{T}_{\mathsf{\dd}} \rho_\kp=\frac{\Phi_\delta(\kp)}{\Phi_\delta(-\kp)}\rho_\kp\    ,
\ee
as well as \eqref{eq:Omega-CKP-dd} and \eqref{eq:Urel-CKP}, we derive
\be \label{eq:bukrel-CKP} 
\wt{\buk}   =\left(\boldsymbol h(\wt \xi,\dd,\dd) \wt{\bU_\xi}  \boldsymbol h^{-1}(\wt\xi,\delta,\tLd)\bO -\boldsymbol h(\wt\xi,\delta,\Ld)\right)\boldsymbol h^{-1}(\xi,-\delta,\Ld)\buk  \   . 
 \ee
Taking the central element of the above equation, and setting $\phi=\tee\,\buk $\,, $\psi_\ar=\tee\, h^{-1}(\xi,\ar,\Ld) \cdot \buk$, we  then have 
\be 
\wt{\phi}=-{\phi} +\boldsymbol h (\xi,-\dd,-\dd)  \wt{v_\dd}\psi_{-\dd} \  ,
\ee
which can be rearranged as
\be \label{eq:psi-d}
\psi_{-\dd}=\frac{1}{\boldsymbol h (\xi,-\dd,-\dd)  \wt{v_\dd}} \left(\wt{\phi}+{\phi}\right ) \   .
\ee
Similar relations hold for other directions, which are
\begin{align}
\psi_{-\ven}=\frac{1}{\boldsymbol h (\xi,-\ven,-\ven)  \wh{v_\ven}}\left (\wh{\phi}+{\phi} \right) \   ,  \label{eq:psi-v} \\
\psi_{-\nu}=\frac{1}{\boldsymbol h (\xi,-\nu,-\nu)  \wb{v_\nu}} \left(\wb{\phi}+{\phi}\right ) \   .
\end{align}
From $\tee\, \boldsymbol h^{-1}(\wt\xi,\ar,\Ld)\cdot \eqref{eq:bukrel-CKP}$ we get
\begin{align}\label{eq:psi}
\boldsymbol h(\wt \xi,\ar,\dd)\wt{\psi_{\ar}}=-\boldsymbol h(\xi,\ar,-\dd){\psi}_{\ar} +\boldsymbol h(\xi,-\dd,-\dd)\left(1-\boldsymbol h(\wt \xi,\ar,\dd) \wt{s_{\ar,\dd}}\right){\psi}_{-\dd}  \ .  
\end{align}
Putting $\ar=-\ven$ in \eqref{eq:psi}, then from relations \eqref{eq:psi-d}, \eqref{eq:psi-v} and \eqref{eq:psi-v} $\wt{}$ , we obtain the Lax triplet of the lattice CKP equation
\begin{align}
\frac{\boldsymbol h(\wt \xi,\dd,-\ven)}{\boldsymbol h(\wt \xi,-\ven,-\ven)}\left(\wt \phi+\wt {\wh\phi} \right)=-\frac{\boldsymbol h(\xi,-\dd,-\ven)}{\boldsymbol h( \xi,-\ven,-\ven)}\cdot\frac{\wt{\wh{v_\ven}}}{\wh{v_\ven}}\left(\phi+\wh \phi\right) +\left(1- \boldsymbol h(\wt \xi,\dd,-\ven )\wt{s^{}_{-\ven,\dd}}\right)\frac{\wt{\wh{v_\ven}}}{\wt{v_\ven}}\left(\phi+\wt \phi\right) \ , \nn
\end{align}
and the other two similar linear equations if we interchange $\{\,\wt{}\ ,  \dd\}\leftrightarrow \{\,\wh{}\ ,  \ven\} \leftrightarrow\{\,\bar{} \ ,  \nu\}$. Variables like $v_\ven$ and $s^{}_{-\ven,\dd}$ can be replaced by the $\tau$ or $f$-function, so that the compatibility condition of the Lax triplet yields the lattice CKP equation.

\section{Elliptic soliton solution structure}
\setcounter{equation}{0}

It is clear that exact solutions of the CKP equation \eqref{eq:CKP-tau-f} can be expressed by the explicit structure of the form $f=\sg(\xi) \det\left({\boldsymbol 1}+\bOm_\xi\cdot\bC\right)$.  We now point out a special solution that is the elliptic multi-soliton solution for the lattice CKP equation. The key idea is to choose a particular measure in \eqref{eq:inteq} written as 
\be \label{measure-CKP}
{\rm d}\mu(\ell ,\ell ^\prime)=\left(\frac{1}{2\pi  \mathrm{i}}\right)^2 \sum_{i,j=1}^N A_{i,j}
 \frac{{\rm d}\ell~ {\rm d}\ell'}{(\ell- \kappa_{i})(\ell'- \kappa_{j})} \  ,
\ee
where $A_{i,j}=A_{j,i}$ are constants, and $\mathrm{i}$ is the imaginary unit. In this sense, the measure exhibits symmetry, i.e., ${\rm d}\mu(\ell ^\prime, \ell)= {\rm d}\mu(\ell ,\ell ^\prime)$. Therefore, the linear integral equation \eqref{eq:inteqa} is reduced to 
\begin{equation}\label{eq:uksoliton-CKP}
\bu_\kappa +\sum_{i,j=1}^N A_{i,j} \,\rho_\kappa\, \Phi_{\xi}(\kappa+ \kappa_{j} )\, \sg_{\kappa_{j}} \bu_{\kappa_{i}}=\rho_\kappa{\Phi_\xi(\Ld)}  \bc_\kappa \ .
\end{equation}

Putting $\kappa=\kappa_s$, where $s=1,2,\dots, N$ in the above equation \eqref{eq:uksoliton-CKP}, then from equation \eqref{eq:uksoliton-CKP}, we have a linear
system for the quantities $u_{\kappa_s}$ :
\begin{align}
  \left(\bu_{\kappa_1},\bu_{\kappa_2},\dots,\bu_{\kappa_N}\right) ({\bI_N}+\bA\bM) = \left(\bc_{\kappa_1},\bc_{\kappa_2},\dots,\bc_{\kappa_{N}}\right)
 \, \bbR  \ , \label{eq:uksoliton-sq-CKP}
\end{align}
in which
\begin{align}
 \bbR=  {\rm diag} \Big(\rho_{\kappa_1} \Phi_\xi(\kappa_1),\rho_{\kappa_{2}}\Phi_\xi(\kappa_{2}),\dots,\rho_{\kappa_{N}}\Phi_\xi(\kappa_{N})\Big)     \  ,
\end{align}
$\bA =(A_{i,j} )_{N\times N}$, $\bI_N$ is an identity matrix of size $N$ and $\bM$ is an $N \times N$ matrix with elements
\begin{equation}\label{eq:MCauchy-CKP}
M_{j,i}=
\rho_{\kappa_{i}} \Phi_{\xi}(\kappa_{i}+ \kappa_{j}) \sg_{\kappa_{j}} \  .
\end{equation}
The discrete plane wave factors are given by $$\rho_{\kappa_{i}} =\sg_{\kappa_{i}}=\left(\frac{\Phi_{\dd}(\kappa_i)}{\Phi_{\dd}(-\kappa_i)} \right)^n\left(\frac{\Phi_{\ven}(\kappa_i)}{\Phi_{\ven}(-\kappa_i)}\right)^m\left(\frac{\Phi_{\nu}(\kappa_i)}{\Phi_{\nu}(-\kappa_i)}\right)^l\rho_i^0 \ , \quad (i=1,\dots,N) \  .$$

The explicit elliptic soliton solutions can now be obtained by assuming appropriate values for the coefficients $A_{i,j}$, so that the matrix ${\bI_{N}}+\bA \bM$ becomes invertible. So in this case, solutions to \eqref{eq:uksoliton-CKP} or \eqref{eq:uksoliton-sq-CKP} could be written out explicitly. 
Making use of \eqref{eq:U-in} and \eqref{measure-CKP} we can also obtain the explicit expression for $\bU_\xi $, rewritten in the form
\begin{align*}
  {\boldsymbol U}_\xi = \left(\bc_{\kappa_1},\bc_{\kappa_2},\dots,\bc_{\kappa_{N}}\right)
 \, \bbR  \, (\bI_{N}+\bA\bM) ^{-1} \bA \, \bbR \,  \,{}^{t}\left(\bc_{\kappa_1},\bc_{\kappa_2},\dots,\bc_{\kappa_{N}}\right)  \ ,
\end{align*}
and $u=\brr  \, (\bI_{N}+\bA\bM) ^{-1} \bA \, \,  \,{}^{t}\brr  \ ,$
with $\brr=\Big(\rho_{\kappa_1} \Phi_\xi(\kappa_1),\rho_{\kappa_{2}}\Phi_\xi(\kappa_{2}),\dots,\rho_{\kappa_{N}}\Phi_\xi(\kappa_{N})\Big). $

Moreover,  we can write the quantities
introduced in \eqref{eq:tau} and \eqref{eq:vws} in the following way: 
\begin{align}\label{eq:tausol}
&\tau = \det\left({\bI_{N}}+\bA \bM \right)\  ,\\
&v_\ar =w_\ar= 1 - \brr\,\bh^{-1}( \xi,\ar,\bK)\cdot\, (\bI_{N}+\bA\bM) ^{-1} \bA \, \,{}^{t}\brr 
\   ,   \\
& s_{\ar,\bb} = \brr\,\bh^{-1}( \xi,\ar,\bK)\cdot\, (\bI_{N}+\bA\bM) ^{-1} \bA \,\cdot \boldsymbol h^{-1}( \xi,\beta,\bK) \,{}^{t}\brr  \   ,  
\end{align}
in which $\bK={\rm diag} \left(\kappa_1, \kappa_{2},\dots,\kappa_{N}\right).   $ 

The elliptic soliton expression for the $\tau$-function \eqref{eq:tausol} can be written in Hirota form by 
expanding the determinant and computing the matrix invariants from  the 
sums of principal minors of $\bA\bM$ using the Cauchy-Binet theorem, cf. 
\cite{Y-KN}. We will omit the resulting explicit formula here which looks very similar.

\section{Semi-continuous CKP equations}
\setcounter{equation}{0}

The advantage of \eqref{eq:CKP-tau-f} over \eqref{eq:CKP} is that the presence 
of the lattice parameters $\delta,~\ven$ and $\nu$ in the elliptic 
parametrisation allows one to compute continuum limit in a straightforward way, 
other than to rely on guess work. These limits are rather subtle, cf. 
\cite{book-HJN-2016}, and rely on the analysis of how the discrete plane wave factor 
$\rho_\kappa$ is transformed into exponential functions under such limits. 

\paragraph{Straight continuum limit:} 
The discrete plane wave factor is given by $\nu \to 0$, $l\to\infty$ such that $l \nu\to x$, leading to 
$$\rho_{\kappa} =\left(\frac{\Phi_{\dd}(\kappa)}{\Phi_{\dd}(-\kappa)} \right)^n\left(\frac{\Phi_{\ven}(\kappa)}{\Phi_{\ven}(-\kappa)}\right)^m\left(\frac{\Phi_{\nu}(\kappa)}{\Phi_{\nu}(-\kappa)}\right)^l
\quad\to\quad e^{2\zeta(\kp)x}\left(\frac{\Phi_{\dd}(\kappa)}{\Phi_{\dd}(-\kappa)} \right)^n\left(\frac{\Phi_{\ven}(\kappa)}{\Phi_{\ven}(-\kappa)}\right)^m
.$$
While in this limit the variable $\xi=\xi_0-2n\delta-2m\ven-2l\nu\to \xi_0-2x-2n\delta-2m\ven $.  Now we impose this limit on the lattice equation \eqref{eq:CKP-tau-f}. Using the above limit and the expansions in powers of $\nu$, we arrive at the coefficient of the leading term of order $\mathcal{O}(\nu^2)$:
\begin{eqnarray} \label{eq:semi-CKP-f-1}
&&\Big(-\sigma ^2(\delta -\epsilon)\,f \, \partial_x \wh{\wt {f}}   +\sigma^2 (\delta -\epsilon) \Big (4\, (\zeta (\delta )+\zeta (\epsilon ) )\, f +\partial_x f \Big) \wh{\wt { f}} \nonumber \\ 
 && 
\quad \quad\quad +\sigma^2 (\delta +\epsilon) \Big(\wh {f}\, \partial_x \wt {f}- \left (4\, (\zeta (\delta)-\zeta (\epsilon))\, \wh{f}  +\partial_x \wh {f} \right)\wt {f} \Big)\Big)^2  \\ 
 && 
\quad = \,   4\, \sigma ^2(\delta -\epsilon) \sigma ^2 (\delta +\epsilon ) \Big( (4\,f\, \zeta (\epsilon)+\partial_x f)\,\wh{f}-f \,\partial_x \, \wh {f}\Big) \Big( (4  \wt {f} \,\zeta (\epsilon)+\partial_x  \wt {f})\, \wh{\wt{f}} - \wt {f}\, \partial_x \wh{\wt {f}} \Big)    \  ,\nonumber
\end{eqnarray}
which is the semi-discrete CKP equation.

\paragraph{Skew continuum limit:}
The discrete plane wave factor is given by
$$\rho_{\kappa} =\left(\frac{\Phi_{\dd}(\kappa)}{\Phi_{\dd}(-\kappa)} \right)^N\left(\frac{\Phi_{\ven}(\kappa)}{\Phi_{\ven}(-\kappa)}\right)^m\left(\frac{\Phi_{\nu}(\kappa) \Phi_{\dd}(-\kappa)}{\Phi_{\nu}(-\kappa) \Phi_{\dd}(\kappa)}  \right)^l,$$
with the change of independent variables $(n,m,l) \to (N=n+l,m,l)$. By setting $q=\nu-\dd$ and performing the limit  $n\to-\infty, ~l\to \infty ,~q \to 0$ such that $N$ fixed and $ql=x$ finite, we have its skew continuum limit 
$$ \rho_{\kappa}  \quad \to \quad  e^{\zeta(\kp+\dd)x+\zeta(\kp-\dd)x}\left(\frac{\Phi_{\dd}(\kappa)}{\Phi_{\dd}(-\kappa)} \right)^N\left(\frac{\Phi_{\ven}(\kappa)}{\Phi_{\ven}(-\kappa)}\right)^m
.$$
While in this limit the variable $\xi=\xi_0-2n\delta-2m\ven-2l\nu\to \xi_0-2x-2N\dd-2m\ven $. 
Using the above shew limit and the expansions in powers of $q$, we arrive at the coefficient of the leading term of order $\mathcal{O}(q^2)$:
\begin{eqnarray} \label{eq:semi-CKP-f-2}
 &&\sigma ^2(2 \delta) \sigma ^2(\delta -\epsilon) \sigma^2 (\delta +\epsilon) \Big(f_{N+1,m+1} (2 f_{N+1,m} \left (\zeta (\delta +\epsilon)-\zeta (\delta -\epsilon)\right)+\partial_x {f}_{N+1,m})\nonumber
 \\ 
 && 
\quad  - f_{N+1,m} \partial_x {f}_{N+1,m+1}\Big)^2=  4  \,\Big(f_{N+1,m} f_{N+2,m+1}\sigma ^2(\delta -\epsilon)  - f_{N+1,m+1} f_{N+2,m} \sigma^2 (\delta +\epsilon)\Big)    \nonumber \\ 
 &&
\quad  \quad  \times \Big(f_{N,m}f_{N+1,m+1}\sigma^2 (\delta -\epsilon)-f_{N,m+1} f_{N+1,m} \sigma^2 (\delta +\epsilon) \Big)  \ ,
\end{eqnarray}
which is another semi-discrete CKP equation. A non-autonomous differential-difference version of the CKP equation was presented in \cite{2020-sdKP-FN}. Fully continuous $\mathbf{C}$-type 
differential equations arising from the direct linearisation scheme were presented in 
\cite{2017-DL-FN-KP}.

\section{Conclusions and discussions}
\setcounter{equation}{0}

In this paper, we have set up a DL scheme for elliptic solutions of the CKP equation, 
based on integral equations with the elliptic Cauchy kernel. From the scheme we constructed  
elliptic $N$-soliton solutions of the lattice CKP equation, but in principle it also 
provides more generally to inverse scattering type solutions. 
Furthermore, the DL structure not only yields exact solutions, but also 
establishes relationships between the various equations in the scheme, i.e. 
equations for the quantities $u$, $v_\alpha$ and $s_{\alpha,\beta}$. 
Furthermore, as already shown in \cite{2017-DL-FN-dKP,2020-sdKP-FN}, it 
helps to unify the various branches of the discrete and semi-discrete 
KP systems. 

The fully discrete KP family of equations is often regarded as one of the most fundamental 
integrable systems, since many discrete and continuous integrable systems in 1+1 dimensions and 
1-dimensional systems, such as the Toda chain and Calogero-Moser type systems of ODEs can be 
obtained as dimensional reductions. In particular elliptic pole-type solutions can be obtained 
from the lattice KP systems, cf. \cite{1996-NijKuzRagn-dtRuijs,1997-KLWZ-dKP elliptic,2019-RZ-dynamicsBKP}, and \cite{2019-Zabr-ellsols} for a review.

As an outgrowth of the results in this paper the following directions of 
progress can be considered:
to develop the generating PDE for the CKP system, i.e. the associated 
(system of) continuous equations in terms of the Miwa variables; 
the construction of pole-type solutions from the straight continuum limit 
of the CKP equation leading to elliptic multi-particle systems, cf. e.g. \cite{1997-KLWZ-dKP elliptic}, 
and a systematic study of reductions to 1+1-dimensional equations by 
imposing further conditions on the measures ${\rm d}\mu(\ell,\ell^\prime)$. 
Also ahead lie the question about the construction of algebro-geometric solutions, and how to  extend the results to the DKP equation, cf. \cite{2022-eKP-FN}.

\vskip 15pt \noindent{\bf Acknowledgments}~
\vskip 7pt
This project is supported by the NSF of China (Nos.12271334, 12326428, 12471235).
 
\section*{Appendix: derivation of the identity \eqref{eq:hh-identity}}
\def\theequation{A.\arabic{equation}}
\setcounter{equation}{0} 

To derive equation \eqref{eq:hh-identity} we use the definition of $\bh$ in order to write 
\begin{align} 
& \frac{\boldsymbol h(\wt{\xi},\delta,\lambda)}{\boldsymbol h(\wt{\xi},\alpha,\lambda)\,\boldsymbol h(\xi,-\delta,\lambda)} 
=\frac{\Phi_{\wt{\xi}+\ar}(\ld)}{\cancel{\Phi_{\wt{\xi}}(\ld)}\,\Phi_\ar(\ld)}\,
\frac{\cancel{\Phi_{\wt{\xi}}(\ld)}\,\Phi_\delta(\ld)}{\cancel{\Phi_{\wt{\xi}+\delta}(\ld)}}
\,\frac{\cancel{\Phi_{\xi-\delta}(\ld)}}{\Phi_\xi(\ld)\,\Phi_{-\delta}(\ld)}\ \nonumber \\
& =\frac{\Phi_{\wt{\xi}+\ar}(\ld)\,\Phi_\delta(\ld)}{\Phi_\alpha(\ld)\,\Phi_\xi(\ld)\,\Phi_{-\delta}(\ld)}= 
\frac{\Phi_{\wt{\xi}}(\alpha+\ld)\,\Phi_\delta(\ld)}{\Phi_{\wt{\xi}}(\alpha)\,\Phi_\xi(\ld)\,\Phi_{-\delta}(\ld)}\ , \label{eq:PhiPhi}
\end{align}
noting that $\wt{\xi}+\delta=\xi-\delta$, and 
where in the last step use has been made of the relation 
\[ \frac{\Phi_\ar(x)}{\Phi_\beta(x)}= \frac{\Phi_{\ar-\beta}(\beta+x)}{\Phi_{\ar-\beta}(\beta)}\ , \]
which follows directly from the definition of the function $\Phi$.  
We want to write \eqref{eq:PhiPhi} in the form 
\be\label{eq:AB} 
\frac{A}{\bh(\xi,\ar,\ld)}-\frac{B}{\bh(\xi,-\delta,\ld)}= 
\frac{A\,\Phi_\xi(\ar+\ld)\,\Phi_\xi(-\delta)-B\Phi_\xi(\ld-\delta)\,\Phi_\xi(\ar) }{\Phi_\xi(\ar)\,\Phi_\xi(\ld)\,\Phi_\xi(-\delta)}\ , 
\ee 
with $A$, $B$ independent of $\lambda$. Comparing \eqref{eq:PhiPhi} and \eqref{eq:AB}, and ignoring the common factor $\Phi_\xi(\ld)$ in the denominators, 
we can compute the remaining factor in \eqref{eq:PhiPhi} as
\begin{align}
& \frac{\Phi_{\wt{\xi}}(\alpha+\ld)\,\Phi_\delta(\ld)}{\Phi_{\wt{\xi}}(\alpha)\,\Phi_{-\delta}(\ld)}=-\frac{\sg(2\delta)}{\Phi_{\wt{\xi}}(\ar)}
\Phi_{\wt{\xi}}(\ar+\ld)\,\Phi_{2\delta}(\ld-\delta)= \nonumber \\ 
&= -\frac{\sg(2\delta)}{\Phi_{\wt{\xi}}(\ar)}\left[\Phi_\xi(\ar+\ld)\,\Phi_{2\delta}(-\ar-\delta)+\Phi_{\wt{\xi}}(\ar+\delta)\,\Phi_\xi(\ld-\delta)\right] \ . \label{eq:Phisum}
\end{align}
In the second step we have used the identity 
\[ \Phi_\alpha(x)\,\Phi_\beta(y)=\Phi_{\alpha+\beta}(x)\,\Phi_\beta(y-x)+ 
\Phi_\alpha(x-y)\,\Phi_{\alpha+\beta}(y)\  ,  \] 
which is a consequence of the well-known 3-term addition formula for the $\sg$-function. 
The two terms of \eqref{eq:Phisum} compare with the $A$ and $B$ terms in \eqref{eq:AB} 
respectively. Thus, we can identify 
\[ \frac{A}{\Phi_\xi(\ar)}=-\frac{\sg(2\delta)}{\Phi_{\wt{\xi}}(\alpha)}\,\Phi_{2\delta}(-\ar-\delta)\  , 
\quad \frac{B}{\Phi_\xi(-\delta)}=\frac{\sg(2\delta)}{\Phi_{\wt{\xi}}(\alpha)}\,\Phi_{\wt{\xi}}(\ar+\delta)\  ,  \]  
which after writing these out in terms of $\sg$ functions, yields 
\[ A= \frac{\bh(\xi,\ar,-\delta)}{\bh(\wt{\xi},\ar,\delta)} , \quad B= \frac{\bh(\xi,-\delta,-\delta)}{\bh(\wt{\xi},\ar,\delta)}\  . \] 
Inserting these into \eqref{eq:AB} and equating this to the first expression in \eqref{eq:PhiPhi}, we obtain the identity \eqref{eq:hh-identity}. The analogous identity \eqref{eq:hhh-identity} can be proven along a similar line, or simply by changing $\delta\to -\delta$ and reversing the shifts, while replacing $\ar$ by $\beta$.

{\small


\begin{thebibliography}{99}

\bibitem{Akh} N.I. Akhiezer,
{ Elements of the Theory of Elliptic Functions}, Translation of Mathematical Monographs, Vol. { 79},
(AMS, Providence RI, 1990).

\bibitem{Atkinson-Nij2010} J. Atkinson, F.W. Nijhoff, A constructive approach to the soliton solutions of integrable quadrilateral lattice equations, Commun. Math. Phys., 2010, 299: 283-304. 

\bibitem{2015-Bobenko-CKP} A.I. Bobenko, W.K. Schief, Discrete line complexes and integrable evolution of minors, Proc. R. Soc. London A, 2015,  471: 20140819 (23pp).

\bibitem{2017-Bobenko-CKP}  A.I. Bobenko, W.K. Schief,  Circle complexes and the discrete CKP equation, Int. Math. Res. Notices, 2017, 5: 1504-1561.

\bibitem{2002-Bobenko-MDC} A.I. Bobenko, Yu.B. Suris, Integrable systems on quad-graphs, Int. 
Math. Res. Notices, 2002, 11: 573-611.

\bibitem{Cayley} 
A. Cayley, On the theory of linear transformations, Cambridge Math. J., 1845, 4: 193-209.

\bibitem{2018-Chang}  X.K. Chang, X.B. Hu, S.H. Li, Degasperis-Procesi peakon dynamical system and finite Toda lattice of CKP type, Nonlinearity, 2018, 31: 4746-4775.

\bibitem{2020-Chang} 
X.K. Chang, S.H. Li, S. Tsujimoto, G.F. Yu,
Two-parameter generalisations of Cauchy bi-orthogonal
polynomials and integrable lattices, ArXiv: 2007.05998 (2020). 

\bibitem{2025-Chang} 
S.H. Li, S. Tsujimoto, R. Watanabe, G.F. Yu,
Cauchy-Jacobi orthogonal polynomials and the discrete CKP equation, ArXiv: 2504.18891 (2025). 

\bibitem{2010-Doliwa-CKP} A. Doliwa, The C-(symmetric) quadrilateral lattice, its transformations and the
algebro-geometric construction, J. Geom. Phys., 2010, 60: 690-707.




\bibitem{2017-DL-FN-dKP} W. Fu, F.W. Nijhoff, Direct linearizing transform for three-dimensional discrete integrable systems: the lattice AKP, BKP and CKP equations, Proc. R. Soc. A, 2017, 473(2203): 20160915 (22pp).

\bibitem{2017-DL-FN-KP} W. Fu, F.W. Nijhoff, Linear integral equations, infinite matrices, and soliton hierarchies, J. Math. Phys., 2018, 59(7): 071101 (28pp).

\bibitem{2020-sdKP-FN} W. Fu, F.W. Nijhoff, On non-autonomous differential-difference AKP, BKP 
and CKP equations, Proc. Roy. Soc. A, 2021, 477(2245): 20200717 (20pp).




\bibitem{2022-eKP-FN} W. Fu, F.W. Nijhoff, On a coupled Kadomtsev-Petviashvili system associated with an elliptic curve, Stud. Appl. Math., 2022, 149(4): 1086-1122.

\bibitem{book-HJN-2016} J. Hietarinta, N. Joshi, F.W. Nijhoff, 
Discrete Systems and Integrability, Cambridge University Press, 2016.  

\bibitem{2007-Holtz-CKP}  O. Holtz, B. Sturmfels, Hyperdeterminantal relations among symmetric principal minors, J. Algebra, 2007, 316(2): 634-648.

\bibitem{1980-JM}  
M. Jimbo, T. Miwa, Solitons and infinite dimensional Lie algebras, Publ. RIMS, 1983, 19: 943-1001. 

\bibitem{1996-Kashaev-CKP} R. Kashaev, On discrete three-dimensional equations associated with the local Yang-Baxter equation, Lett. Math. Phys., 1996, 33: 389-397.

\bibitem{2016-Kenyon-CKP} R. Kenyon, R. Pemantle, Double-dimers, the Ising model and the hexahedron recurrence, J. Combin. Theory Ser.A, 2016, 137: 27-63.




\bibitem{1997-KLWZ-dKP elliptic} I.M. Krichever, O. Lipan, P. Wiegmann,  A. Zabrodin, Quantum integrable models and discrete classical  Hirota equations, Commun. Math. Phys., 1997, 188: 267-304. 


\bibitem{2022-JNS-ellKdV} X. Li, D.J. Zhang, Elliptic soliton solutions: tau functions, vertex operators and bilinear identities, J. Nonlinear Sci., 2022,  32(5): No.70 (53pp).

\bibitem{1989-MailNij-localYB}
J-M. Maillet, F.W. Nijhoff, Integrability for multidimensional lattice
models, Phys. Lett. B, 1989, 224: 389-396.

\bibitem{2010-Nijhoff-ellABS} 
F.W. Nijhoff, J. Atkinson,  Elliptic $N$-soliton solutions of ABS lattice equations, Int. Math. Res. Notices, 2010, 2010(20): 3837-3895.

\bibitem{1983-DL-NQC} F.W. Nijhoff, G.R.W. Quispel, H.W. Capel,
	Direct linearization of nonlinear difference-difference equations,	                             Phys. Lett. A, 1983, 97:125-128.

\bibitem{1996-NijKuzRagn-dtRuijs} 
F.W. Nijhoff, O. Ragnisco, V. Kuznetsov, Integrable time-discretization of the
Ruijsenaars-Schneider model, Commun. Math. Phys., 1996, 176: 681-700.

\bibitem{2019-ellBSQ} F.W. Nijhoff, Y.Y. Sun, D.J. Zhang, Elliptic solutions of Boussinesq type lattice equations and the elliptic $N$th root of unity,  Commun. Math. Phys., 2023, 399: 599-650.

\bibitem{2001-Nijhoff-CAC} F.W. Nijhoff,  A.J. Walker,
                 The discrete and continuous Painlev\'{e} VI hierarchy and the Garnier systems,
                 Glasgow Math. J., 2001, 43A: 109-123.
                 



\bibitem{1994-book} I.M. Gel'fand, M.M. Kapranov, A.V. Zelevinsky, Discriminants, Resultants, and 
Multidimensional Determinants, Mathematics: Theory \& Applications, Birkh\"auser, Boston, 1994.


\bibitem{2019-RZ-dynamicsBKP} 
D. Rudneva, A. Zabrodin, Dynamics of poles of elliptic solutions to the BKP equation, 
ArXiv: 1903.00968. 

\bibitem{2003-Schief-CKP} W.K. Schief, Lattice geometry of the discrete Darboux, KP, BKP and CKP equations. Menelaus’ and Carnot’s theorem, J. Nonlinear Math. Phys., 2003, 10(suppl. 2): 194-208.

\bibitem{2006-Schief-CKP}  W.K. Schief, Application of an incidence theorem for conics: Cauchy problem and integrability of the dCKP
equation, J. Phys. A: Math. Gen., 2006,  39: 1899-1913.




\bibitem{2009-Tsarev} S.P. Tsarev, T. Wolf, Hyperdeterminants as integrable discrete systems, J. Phys. A: Math. Theor., 2009, 42: 454023 (9pp).

\bibitem{Y-KN} S. Yoo-Kong, F.W. Nijhoff, {Elliptic $(N,N')$-soliton solutions of the lattice 
Kadomtsev-Petviashvili equation}, J. Math. Phys., 2013, 54: 043511 (20pp).



\bibitem{2019-Zabr-ellsols}
A. Zabrodin, Elliptic solutions to integrable nonlinear equations
and many-body systems, J. Geom. Phys., 2019, 146: 103506 (26pp). 



\end{thebibliography}
\end{document}